# TupperSats: Thinking Inside the Box for Space Systems Engineering

David Murphy[a]*, Robert Jeffrey[a], Deirdre Coffey[a], Morgan Fraser[a], Sheila McBreen[a], Lorraine Hanlon[a]

[a] *School of Physics, University College Dublin, Belfield, Dublin 4, Ireland.*
\* david.murphy@ucd.ie

**Abstract**
As part of University College Dublin's MSc in Space Science & Technology curriculum, student teams, over the course of a single term, are required to design, build, launch (on a meteorological balloon), operate and recover their own payload on a standardised platform. Each 'TupperSat' must be built from, or contained within, a household plastic storage container, It must weigh less than 1kg, be able to determine location, altitude, internal and external temperature and air pressure, and be able to broadcast this information using an in-house communications system. Students must also design and build a scientific payload or novel technology demonstration to fly on their TupperSat. Notable examples include an earth observation vegetation sensor, particle sample return, gamma-ray detector, and air-bag landing system.
The instructors play the role of customer and launch authority. The students are provided with a number of standard components including a Raspberry Pi single-board computer, a 5000 mAH battery, high-altitude-compatible GPS unit, temperature and pressure sensors and, a small low-power radio transceiver module developed specifically for the course based on the LoRa standard. Teams are given a budget of EUR 100 (≈ $115) to purchase additional materials and to build their payload. The students learn space industry practices by full immersion in a typical space project development cycle. TupperSat design and payload concepts are pitched by the student teams at a Preliminary Design Review; plans are well developed before a Critical Design Review, and the team must pass a Flight Readiness Review before being granted permission to launch. Good project management is crucial in order to meet deadlines and secure a launch at the end of the term.
The module has run for 6 years with the participation of 64 students so far. As it has become more popular and student participation has grown, the module has been modified to allow for more ambitious and challenging projects which further motivates the students.
The module syllabus and student learning outcomes are presented, along with implementation lessons learned.

**Acronyms/Abbreviations**
TupperSat Data Relay Satellite System (TDRSS)
Preliminary Design Review (PDR)
European Cooperation for Space Standardization (ECSS)
Tuppersat Telemetry Thingamabob (T³)

1. **Introduction**
One problem inherent with space projects is the length of time from concept through launch and operations.

The aim of the Satellite Subsystems laboratory is to take students through the complete satellite system development process in a single 12-week term. The goal is to make these small satellites (which we call 'TupperSats') as capable as possible and to (i) develop a payload compatible with platform, budget and mass constraints; (ii) launch these satellites by weather balloon; (iii) operate the satellites and telemeter data and (iv) recover the satellite.

This forms a part of the UCD MSc programme in Space Science & Technology. The motivation for this programme is to equip graduates from engineering or physics backgrounds with technical knowledge regarding the space environment and space applications; detector characterisation, simulation, measurement and performance verification; mission design; space project management and proposal-writing. The curriculum is industry-facing, and was developed in partnership with space companies and agencies, while also building on the research and technical expertise of our academic staff. Importantly, this partnership with industry extends to the delivery of the MSc programme, with leading professionals from the Irish and European space sectors delivering seminars, lectures and workshops throughout the course.

The capstone of the year long programme is a 12-week internship in a space company or agency. The Satellite Subsystems module was designed to prepare students for the internship by exposing them to systems engineering in a practical, rather than classroom setting. The module adopts a tailored version of project phases, from Phase A (Feasibility) to Phase E (Operations), according to the ECSS "System engineering general requirements" standard (ECSS-E-ST-10C).

In this paper, we set out the teaching philosophy and aims underlying the Satellite Subsystems module, describe how it is delivered at UCD, and discuss some of the lessons that we have learnt from our experiences over the 6 years in which we have offered the course. Section 2 provides a brief overview of what a TupperSat is. Section 3 describes the academic and pedagogical goals of the module. Section 4 describes the practical delivery of the module. In Sections 5 & 6, we provide examples of some of the experiments that students have produced, and discuss student experiences and feedback. Finally, in Section 7, we discuss our plans for the class in 2020 and beyond.

2. **TupperSats**

A TupperSat is a small autonomous scientific experiment, designed to fit into a Tupperware-style household container (hence the name, TupperSat) and be launched on a high-altitude balloon. The aim is to encourage student creativity, and so the design requirements for a TupperSat are reasonably lightweight .

A TupperSat must:
- have a mass less than 1kg,
- be built using an enclosure made from a plastic household storage container or similar,
- securely attach to the balloon using an attachment that we specify.

It must be capable of determining, logging and telemetering in real-time its location, altitude, internal temperature, external temperature, air pressure.

The required telemetry must be transmitted in a manner that is compatible with a ground segment which is operated by the instructors, the TupperSat Data Relay Satellite System or TDRSS.

To aid the students in achieving these goals, they are provided with a TupperSat Starter Kit which comprises
a Raspberry Pi model B+ single board computer, a USB GPS module with an `airborne' mode which enables high-altitude use, a TupperSat Telemetry Thingamabob (T³) LoRa transceiver, a DS18B20 1-wire temperature sensor, a waterproof DS18B20 1-wire temperature sensor, an MS5611 I²C pressure sensor, an MCP3008 SPI 8-channel analog to digital converter,
and a USB power bank.

To build a functioning TupperSat, the students need to assemble these components and add their own structure, an antenna, and an experiment payload. They also need to write their own on-board flight software.

The total cost of additional materials for each team is subject to a budget of €100 (≈$115).

3. **Module Design & Conception**

*3.1 Aims & Learning Outcomes*
Students are introduced to project management, project phases, systems engineering, programming for embedded systems using Python and a Raspberry Pi, collaborative tools and version control with Git, and preparing documentation.

*3.1 Context within the MSc Space Science & Technology*
The MSc Space Science & Technology at UCD is a 12-month postgraduate level course designed to equip students with the scientific and engineering skills for a career in the space industry. Students take modules worth a total of 90 ECTS credits, consisting of six 5-credit modules, three 10-credit modules, and an industry placement or research project and dissertation worth 30-credits.

The Satellite Subsystems laboratory is one of the 10-credit modules and occurs in the second term (between January and April). By the time students start the module, they have already completed a 10-credit lab in which they work with UCD's CubeSat simulator, EduCube, [1] which we use to introduce them to satellite components. Alongside the Subsystems lab, they prepare for and attend a Space Mission Design field trip, run jointly with the University of Southampton and Universidad de La Laguna, in which they develop skills in teamwork and project design. The Subsystems laboratory combines elements of both of these modules, with students requiring technical, interpersonal, and organisational skills to succeed.

Students come from a wide range of backgrounds: we require a 2.I or equivalent honours degree in science, engineering, or mathematics, and previous students have come from undergraduate degrees including Physics, Aeronautical Engineering, Electrical Engineering, Mechanical Engineering, Computer Science, and Biomedical Science. Graduates from the programme have gone on to careers in both research and industry, with alumni now working across the Irish and European space sector.

*3.3 Student Assessment*
The student assessment for this module consists of a mix of group and individual submissions, and continuous assessment of students' contributions throughout the module and the launch campaign.

The group submissions are framed around project deliverables: each team produces a report and presentation for the Preliminary Design Review, and prepares a user manual and operations documentation prior to launch. This acts as a formative assessment intended to replicate the major project milestones within the project lifecycle. These group components contribute 25% of the overall module grade.

The individual submission consists of a final project report, in which they summarise the whole design and development process for the project, and analyse the data collected during the flight. This allows us to assess a student's own contribution, and their engagement with and understanding of parts of the project that they might not have worked on directly. This report is worth 15% of the final grade.

The remaining 60% of the grade is awarded by continuous assessment. Students are graded by the instructors throughout the term based on the quality of their effort and their engagement with tasks assigned to them. This includes how well they respond to questions or suggestions from tutors, creativity in problem solving, and quality of team work. It also includes a presentation element, as we ask one student from each team to give short spot talks on their team's progress during the weekly lab. Students also receive credit for their performance during final testing and the launch campaign.

4. **Module Implementation**

*4.1 Project Timeline*
Table 1 briefly summarises the project timeline that we give to the students. The structure is based on the project phasing from the ECSS document "System engineering general requirements" (ECSS-E-ST-10C) [2], but is consolidated to fit the full development and operations cycle into a single term of just over 3 months.

Table 1. Project Timeline

| Project Phase | Week | Task | Deliverable |
|---|---|---|---|
| Proposal | -6 | Students are given the Call for Proposals & their team assignments | |
| | 1 | Come up with payload concept | 3min team pitch |
| | 2 | Develop concept into detailed design and requirements, leading to Preliminary Design Review | 20min team presentation & PDR report incl. design and project management details |
| Production & Qualification | 3-9 | Research, development, prototyping, assembly; Flight Readiness Review | Intermediate code & design deadlines; TupperSat User & Operations manual |
| Operations & Utilisation | 10 | Final pre-flight preparation & testing; Flight Acceptance Review | Final flight-ready TupperSat |
| | 11-12 | Launch window (actual date determined by weather and student schedules) | |
| | 13 | Post-flight analysis and data processing | Final individual report |

Within our simplified project life-cycle, we combine the Feasibility, Preliminary Definition, and Detailed Definition phases of the ECSS scheme into a single Proposal phase. This begins with a Call for Proposals, which is issued to students before the Christmas vacation, and concludes with a Preliminary Design Review at the end of the second week of term. This is followed by a Production & Qualification phase, during which students perform research & development, and build their TupperSat, leading to Flight Readiness Review. Finally, the Operations & Utilisation phase consists of final flight acceptance, launch & recovery, and post-flight analysis.

The main project deadlines consist of a series of reviews: Preliminary Design Review (PDR), Flight Readiness Review, and Flight Acceptance Review, followed by the flight and final report. As noted in Section 3.3, these are tied to the student assessment.

4.1.1 Proposal Phase
In the Proposal Phase, students work closely with the instructors as they refine their ideas into a project that is practical and feasible in the time available. Often, the ideas that teams pitch in their initial presentations are overly ambitious in their technical complexity or scope, and there is then an iterative development process as the instructors encourage them to tighten up their mission objectives and focus on essential, achievable goals.

4.2.2 Production and Operations Phases
Throughout the Production Phase, teams are responsible for the project management and scheduling of their work. The instructors monitor this through regular conversations with the team members, providing technical support and intervening if necessary to help a team put itself back on track.

In the final part of the project, teams perform final testing and launch their experiments. This launch campaign is described in Section 4.4.

*4.2 Project Costs & Staffing*

*4.2.1 Costs*
An approximate costing for the annual consumable expenses of the module (i.e., not including staffing or facilities) is given in Appendix A, based on our experiences over the last 6 years. Typically the module costs run to €2000-€2500 of which the largest single expense (about half the total expenditure) is the launch campaign. This includes the cost of the balloons and helium, and the cost of transportation to the launch site in Co. Fermanagh.

The main factor in the cost scaling is the number of teams: this sets the number of balloons needed for launch. Typically, we have between 12 and 16 students, who we divide into 3 or 4 groups, which means that we only need to launch 2 balloons.

*4.2.2 Staff Allocation*
A module of this nature requires a substantial staffing commitment. This is especially true given that for many of the students (especially those not from an engineering background), this may well be their first encounter with a microcontroller or manufacturing toolkit.

At UCD, the module is run by two members of staff, (a module coordinator and an instructor) who are responsible for the administration, grading, and supervision of students. They typically spend one full day each week in the laboratory with the students to offer advice or help troubleshooting, but they are available at other times throughout the week to answer questions, offer suggestions and monitor the progress of the teams. There is also some time involved in preparing hardware (e.g., the $T^3$). Between the two members of staff, the time commitment to the module is about 4-6 full days per week.

Students are also able to ask for help from other experts within the UCD Space Science group and the School of Physics, and especially those staff members with previous involvement in the module. This might include technical support with their payloads or general advice on their design.

We are also able to draw on the expertise of the Mechanical and Electronics workshops within the School of Physics. Students often consult with these technicians for advice on circuit design, and for help in machining parts for their TupperSat structure or antenna.

*4.3 Supporting Materials*

*4.3.1 Manufacturing*
We provide students with basic tools to build the structure and electronic parts of their TupperSat. We give students access to a 3D printer within the laboratory. This is treated as a time-limited resource, with teams given one day per week to use the printer, unless they can make a convincing case for additional time allocation.

Students are encouraged to make their designs and manufacturing requirements as simple as possible. In most cases, they will build the structure from 3D printed plastic with metal support bars, and with only the antenna requiring outside help from the workshops. This is to make the engineering requirements more accessible to the majority of students who do not come from a mechanical background.

*4.3.2 Programming in Python*
We require the students to write their own onboard software for the Raspberry Pi on their TupperSat, and provide support to the students to do this in Python. We introduce them to concepts like threading, serial IO, and logging and exception handling that are usually unfamiliar even to students who arrive with some experience in Python for data analysis. We provide the teams with a number of code examples and exercises that they must work through to help them build up to the point at which they can write their own complete control program.

*4.3.3 Communications*
The communication system used by TupperSats underwent a number of changes in the early years of the course. It seems that a stable solution to the problem has now been found and has been used reliably for the past three launch campaigns.

In 2014, the students were provided with CanSat radios which were produced by T-minus Engineering for ESA. The radios offered both TTL serial and USB interfaces and were therefore quite easy for students integrate into their TupperSats as well as to use with a computer to act as a ground station. The T-minus radios operated on the 433 MHz ISM band.

When TupperSat1 launched in 2015 with both a T-minus radio and its XBee PRO 868 communication payload, it became quite apparent that the XBee offered a far more reliable quality of service (QoS). The XBee Pro was then adopted as the standard communications system for the 2016 class. From a hardware point of view, the XBee was easy to integrate into the TupperSats, again using a TTL serial interface. To ensure good QoS, the XBee's API mode was required which was more difficult to program than the transparent mode. As communication is mission-critical, a Python library, 'satradio', was written by the instructors, easing the burden of interfacing with the XBee's API mode for the students. The API mode does however offer an excellent introduction to the difference between communication packets and communication frames.

Unfortunately, the XBee PRO 868 was discontinued in 2016 and no stock was available to replenish lost units so an alternative was found in the LoRa radios that were becoming widespread. The current incarnation of the TupperSat communications system is the 'TupperSat Telemetry Thingamabob', or $T^3$. For ease of use and integration, the $T^3$ is a small self-contained unit featuring an RFM95 LoRa radio and an ATMega328 processor which manages the radio interface and communicates with a TupperSat over a TTL serial interface. The $T^3$ features an SMA connection for attachment of an antenna and 6-pin data and power interface for the TupperSats which is compatible with the FTDI serial interface standard pinout. The processor on the $T^3$ runs a slightly customised version of the RadioHead RH_Serial library [3] for message framing, where we have added a signal strength indicator byte to the frame header. This also required us to update our satradio Python API, which we provide to students to interface with the $T^3$, to support the new framing used by the RadioHead library.

The $T^3$ has proved to be reliable and versatile. As it is based on the LoRa standard which is supported by multiple vendors, supply of units should not be an issue. It should therefore form the basis of the TupperSat communications system for many years to come.

*4.3.4 Ground Station & Tracking*

During the first launch campaign in 2015, the students were responsible for operating their own ground station. This first receiver system was very rudimentary, comprising a laptop with a T-minus transceiver and a directional Yagi antenna. The ground software was practically non-existent, consisting simply of a serial terminal displaying the received telemetry packets which luckily were in a reasonably human-readable format.

Despite the challenging conditions, the team was able to track the TupperSat for most of the flight, though very large telemetry gaps existed, particularly at the apex of the flight. The team reestablished communication during the descent but due to the telemetry gap, was not well positioned for recovery and once again lost communication as the TupperSat fell below the horizon in mountainous terrain. The team attempted a search near the last known coordinates which were at just a few hundred metres altitude. Unfortunately poor weather conditions, difficult terrain and fading light led to the search being called off.

The following year, we decided that the instructors should be responsible for providing the ground segment, with students building compatible TupperSats. This led to the creation of the TupperSat Data Relay Satellite System (TDRSS).

The TupperSat Data Relay Satellite System consists of a $T^3$ radio, an off-the-shelf omni-directional 868MHz antenna, and a Raspberry Pi with an internet connection. During the launch campaign, we deploy several (typically 2-3) groundstations, mounted on independent pursuit vehicles. Each ground station runs a Tornado web server that receives and logs telemetry and data packets, and then re-transmits them over HTTP to a central server at UCD.

The central server provides an online web interface, the TupperSat Tracking System, which allows us to monitor the housekeeping telemetry from each satellite, and to track their positions on an online map.

The central server also logs and stores every packet it receives from the TupperSats via all groundstations. We deliver these logged packets to the student teams for their post-flight analysis of their satellite and experimental data. If we have been unable to recover a balloon and its payloads, these centrally stored packets may be the only data that students have to complete their analysis.

*4.4 Launch Campaign*
*4.4.1 Launch Site*
Our launch site is determined by logistical costs, air traffic restrictions and the need to recover the payloads for post-flight analysis. We have used 3 different launch sites.

The first launch occurred in 2015 from the Valentia Observatory in Co. Kerry, operated by Met Eireann. This is the only location within the Irish Aviation Authority (IAA) jurisdiction from which the launch of high altitude balloons is permitted. Unfortunately, the probability of recovery from this site is low: balloons launched from here are likely to be carried out to the Atlantic Ocean; if blown inland, the rugged terrain makes pursuit difficult. This was the case in 2015: the balloons and their payloads were lost in Gougane Barra National Park. It is also too far to travel to the launch site, launch, recover and return to Dublin in a single day: this incurs an additional cost.

The Civil Aviation Authority (CAA), which operates jurisdiction over the airspace in Northern Ireland, imposes fewer restrictions on balloon launches. In 2016, we launched from St. Mary's Primary School near Dungannon, Co. Tyrone. Although both balloons were (eventually) recovered, one landed in Lough Neagh, and the other came down in the Belfast urban area.

Since 2017, we have used a site at Lough Navar Forest Viewpoint near Enniskillen in Co. Fermanagh (coordinates 54.4669° N, 7.9063° W). This site offers several advantages. The prevailing winds generally carry the balloons in a north-easterly or easterly direction towards County Tyrone. This takes us away from IAA airspace and away from large bodies of water and built-up areas. The launch position is elevated at a height of 300m but is well sheltered by the forest, allowing us to fill and launch the balloons with sufficient protection from the wind.

*4.4.2 Launch Permissions & Launch Window*
The launch schedule is constrained by the academic term and so must occur in mid to late April. We obtain permission for up to 3 launches within the 2 week launch window from the UK's Civil Aviation Authority (CAA), at least one month in advance of the first planned launch date. The exemption/permission awarded by the CAA requires the flights to stay within UK controlled airspace and for the experiments to descend by parachute at the termination of the flight. Once conditions (both meteorological and academic) look suitable, the CAA must be informed 72 hours in advance of the actual flight date in order to allow time for them to issue a NOTAM (Notice To Airmen) for the day of launch .

*4.4.3 Flight Predictor*
During the launch window, we use the Cambridge University Spaceflight (CUSF) Landing Predictor (predict.habhub.org) to predict the balloon flight path before launch. Typically, this provides reasonably accurate predictions for the next 72-96 hours. This gives us time to inform students of the launch day, confirm transport arrangements, and obtain the NOTAM.

One deficiency of the CUSF predictor is that it does not provide an estimate of the confidence in the prediction. This would help inform our go/no-go decision. We have also experimented with integrating the CUSF landing predictor with our TupperSat Tracking System, to allow us to obtain updated predictions during the flight and pursuit.

*4.4.4 Balloons & Parachutes*
For redundancy, we generally seek to launch two TupperSats, with their two independent communication systems, on a single balloon. In some years, we have also flown our own backup communication system on the balloons, to reduce the risk of mission loss. We use a simple rope rig to attach the two TupperSats to the parachute, with a piece of plastic pipe to keep the payloads apart. This is shown in Figure 1. This piping serves an additional purpose during launch, where it is acts as a hold-point to enable controlled release of the balloon. Since each TupperSat has a mass of 1kg, the total payload mass (including rigging) is between 3-4 kg.

We use a Totex TA-1000 or TA-1200 balloon supplied by Random Engineering and filled with helium gas to provide lift. The launch pressure is estimated so as to achieve a burst altitude between 25km and 35km; this can be done using the data sheet or an online calculator [4]. Depending on weather conditions and the payload mass, we have found that it can be wise to slightly over-pressurise the balloons, sacrificing final altitude and flight time for a higher free lift, meaning the balloon rises faster, bursts earlier, and travels a shorter distance over the ground. This can make the pursuit and recovery of balloons easier.

For descent, the parachute used is the 48 inch LT Spherachute, both items supplied by Random Engineering. For our typical payloads, this gives a final descent rate of about 7m/s.

The parachute serves as the connection point between the payloads and the balloons.  For security, for any object below the parachute, we require at least two independent connections to the parachute itself.

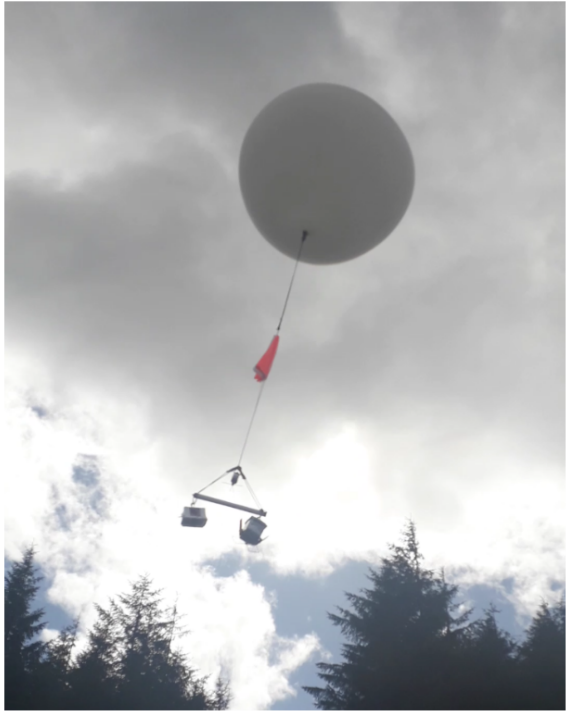

Figure 1. The launch of CrabSat and ABCDSat (2018), showing the balloon, parachute, support rigging & TupperSats. This balloon landed at sea off the Co. Antrim coast.

### 4.4.5 Flight Profile

Figure 2 shows telemetry received from TupperSats flown during the 2018 launch campaign. Both balloons reached altitudes exceeding 24km, before burst and descent. The different ascent rates and flight durations reflect the difficulty in precisely filling the balloons before launch. The balloon carrying CrabSat was underpressured, so rose more slowly and burst at a higher final altitude, while the higher-pressured balloon carrying DiscoSat rose more rapidly and burst sooner. The total flight duration of 3hrs 20mins is the longest flight we have seen. More typically, balloons remain airborne for about 90-120min.

The external temperatures measured by the two balloons drop to about -40°C, while the insulated internal temperature, warmed slightly by the electronics, drops to about 0°C. The pressure measured drops as low as 20 millibar : the MS5611 pressure sensor is one of the few cheap sensors that can tolerate these low pressures.

The weather on the launch day was marginal, with strong winds up to about 20km. Combined with the slow rise of the first balloon, this meant that the balloon, and its payloads, were carried all the way from Co. Fermanagh in south-west Northern Ireland to land in the Irish Sea off the Co. Antrim coast.

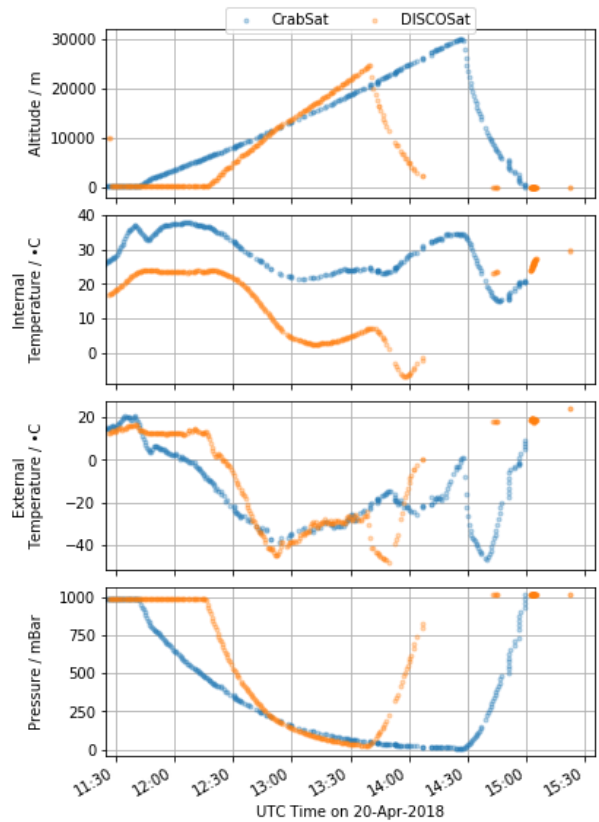

Figure 2. Telemetry received from DiscoSat and CrabSat during the 2018 launch campaign.

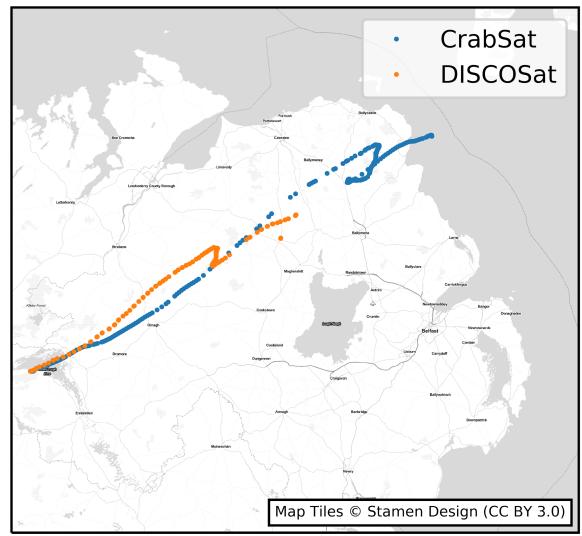

Figure 3. The groundtracks for DiscoSat and CrabSat. DiscoSat was recovered in County Antrim; the balloon carrying CrabSat was lost at sea.

| Year | Mission | Payload | Recovery | Notes |
|---|---|---|---|---|
| 2014 | TupperSat1 | Communications Testbed | Lost in Gougane Barra national park | Launched in 2015 with Can o' Raspberries |
| 2015 | Can o' Raspberries | NDVI land-use sensor | Lost in Gougane Barra national park | |
| 2016 | BetaSat | Geiger-Muller tube | Recovered | |
| | CanSolo | Cellphone | Lost in Lough Neagh | Washed up and recovered 9 months later |
| | HelioSat | Solar tracker | Lost in Lough Neagh | Washed up and recovered 9 months later |
| | MEKSSAT | Spectrometer | Recovered | |
| 2017 | Dustmight Partfinder | Aerogel sample return | Recovered | Data corrupted after accidental activation at low altitude |
| | Pomme de Terre | Sun-tracking solar panel | Recovered | |
| | TupTUPS | Reaction wheel | Recovered | |
| 2018 | ABCDSat | Deployable solar panels | Lost in Irish Sea | |
| | CrabSat | Geiger-Muller tube | Lost in Irish Sea | |
| | DiscoSat | Scintillator Detector | Recovered | |
| | RaccSat | Secondary communications system & image transmission | Recovered | |
| 2019 | AntSat | Experimental airbag landing system | Recovered | Lost radio shortly before launch |
| | CRISP-E | Sun sensor using polarisation | Recovered | |
| | COMUSat | Ultra low pressure $CO_2$ sensor | Recovered | |
| | HotPotato | Infra-red & optical cameras | Recovered | |

Table 2. Past Missions

*4.4.6 Tracking & Pursuit*
For the pursuit, we typically use two to three vehicles equipped with ground stations. Most vehicles start from the launch site immediately after launch, though depending on wind conditions, we usually station an advance car with ground station along the expected trajectory of the balloon, but still able to track the balloons from launch. The pursuit vehicles use the online tracker to keep in close contact with the balloons, with the aim of being within a few hundred metres to at most a couple of kilometres of the landing site.

To get reliable signal from the TupperSats, we need to maintain line-of-sight communication. As soon as this is lost, the low-power radio signal becomes too weak to receive packets. This is not an issue in flight, but becomes difficult at low altitude. Given that we require low transmission cadences (about one packet every 30 seconds) to conserve battery and reduce message collisions, the probability of receiving a valid telemetry packet once the TupperSat is on the ground is small, meaning that the pursuit vehicle must be within a few hundred metres of the balloons when they land.

There are possible solutions to this involving software (e.g., switching to a Landed mode that would provide higher cadence telemetry after landing) or hardware (e.g., using GSM for communication over the internet below a certain altitude). CanSolo in 2016 flew a mobile phone as part of their experiment, in the hope of providing telemetry over internet on the ground: unfortunately, the experiment landed in Lough Neagh.

5.      **Past Missions & Case Studies**

Table 2 lists the missions produced by each cohort since the module was first run in 2014. We have flown 9 balloons carrying 17 satellites, of which only     3 balloons were not recovered: in 2015, 2016 (    later recovered from Lough Neagh), and 2018.

The success of the experiments is more variable, but this is not the principal aim of the exercise. Even in cases where little viable experimental data were obtained, the students still managed to collect telemetry, and gained an understanding of working within a more formal project management structure.

*5.1 DiscoSat & CrabSat*

Two out of the four teams from the 2018 cohort choose to fly experiments to measure ionising radiation at altitude, but taking very different approaches, reflecting the different prior electronics experience within the two groups    .

The CrabSat detector was based on an open source hardware Geiger counter design which was purchased as a kit [5]. The team assembled the kit without significant difficulty. In addition to an LED and speaker to indicate activity, the counter had a TTL serial interface which reported count rates once per second. The team were therefore able to integrate the detector into their TupperSat quite easily to enable logging.

On the other hand, DiscoSat built their own scintillator detector using a LYSO scintillator crystal purchased on eBay, a 3mm SensL SiPM borrowed from the Space Science group at UCD, and their own custom built amplification and peak detection circuit adapted from  [6] and  connected to a microcontroller.

The system worked reasonably well in a counting rate mode, though in the limited time available to them, the team was unable to find a solution to the electronic noise that limited its energy resolution. Nevertheless, this was an extremely impressive result for a radiation detector conceived and built in just a few weeks.

During the balloon flight, the DiscoSat radiation instrument recorded an increase in ionising radiation well correlated with the expected rate due to altitude. The results clearly demonstrated the Pfotzer maximum, a maximum radiation rate at an altitude of approximately 20 km due to particle showers initiated by cosmic rays.

The differing approaches taken by these two teams show that this module can work for students with a wide range of technical experiences. Here, the students involved in CrabSat had less of a background in electronics (they mostly came from undergraduate degrees in physics or mechanical engineering), while the DiscoSat team featured students who did have this experience.

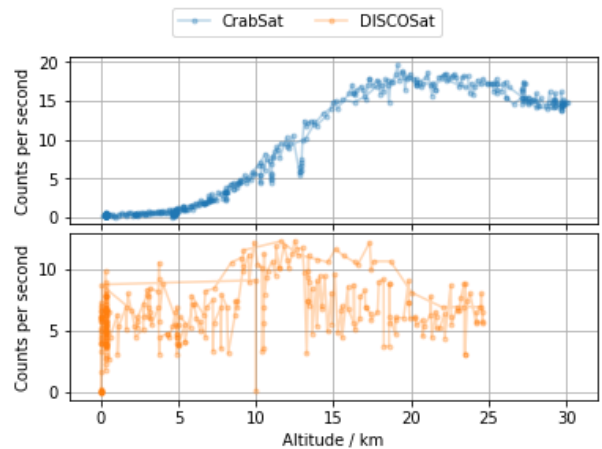

Figure 4.     Count rates vs altitude from the Geiger Muller Tube on CrabSat and the experimental LYSO/SiPM detector on DISCOSat. The Pfotzer maximum is clear in the CrabSat data. The data from the experimental DISCOSat detector is less clear, though there is some evidence for variation at around 12 km.

Both teams found solutions to the same problem that suited their own skill level, and so were able to engage with, learn from, and enjoy the project and module. In other words, the module is not dependent on students having a large amount of prior technical knowledge for it to work or for students to enjoy it.

*5.2 AntSat*
The "AntSat" team in 2019 built a TupperSat which incorporated an inflatable air-bag landing system which was designed to both cushion the landing impact and act as a flotation aid in the event of a water landing.

The system comprised a toroidal airbag placed around the lower part of the TupperSat and a $CO_2$ bicycle tire inflator. The airbag was made from silnylon synthetic fabric, stitched into a torus shape and sealed with epoxy resin. The tire inflator was modified by the team to use an inexpensive small low-power DC motor with reduction gearbox to open the valve and release $CO_2$ into the airbag. The system performed well in ground tests and could quickly be reset using standard and widely available $CO_2$ cartridges.

Unfortunately the system failed to deploy during the real landing. The TupperSat was recovered with the airbag uninflated. It is suspected that the valve or gearing mechanism froze due to the low temperatures experienced at high altitude leading to the deployment failure.

Despite the deployment failure, this is a prime example of the opportunities afforded by the TupperSat concept and a common attitude found in the class. Students often seek out projects in areas in which they desire to increase their capabilities. In this case, the students took on a challenging mechanical project though none of them had a lot of previous experience in mechanical engineering.

This is also a good example of a student team attempting to solve a problem previously encountered by past teams, in this case the water landings from 2016 and 2018. Another example of this is the Tuppersat Tumbling and Undulation Prevention System ("TupTUPs") which aimed to stabilise a camera against the spin which was experienced in videos made by previous TupperSats.

6. **Future Developments**
Over the six years we have run this module, it has continued to grow and evolve as we learn new lessons from each project cycle. At the moment, we are looking at two particular changes that we hope to implement over the next two years.

We are looking at adjusting the timetable and deliverables to introduce an explicit Critical Design Review phase. Currently, we require only that students submit a Preliminary Design Review, which is a detailed document that includes descriptions of the technical designs and team project management. However, this is submitted too early in the project for students to have had a chance to do any prototyping, iteration or research & development. This means that the final designs can often bear little resemblance to that described in such detail at PDR. We plan to retitle this detailed report as the CDR, and to move it closer to the midpoint of the project, with the expectation that the designs presented in the CDR would be close to final, and be the result of a significant R&D effort. We would then introduce a reduced PDR as a 20min team presentation and 4-6 page report, summarising the proposed design and the proposed team structure briefly but in concise detail. Our hope is that doing this will make it easier for us to track changes in the teams' designs, and that by giving an additional milestone to students, we will help them to adhere more closely to real-world project lifecycles.

The other change that we want to make this year is around test campaigns before launch. Students rarely appreciate the importance of testing, especially with such limited time and financial resources. We will require students to document their test campaigns with test plans and short test reports, which will contribute to their final grade. To guide them, we will run two mandatory tests: a "fit check" to confirm that their structure can be connected to the launch balloon well in advance of launch date, and an antenna test.

7. **Conclusions**
The Satellite Subsystems module was designed to provide students with a unique opportunity to learn and experience systems engineering and project management in a practical, hands-on, and enjoyable way. The case studies we have described here show that the module offers something to students from a wide range of academic backgrounds, and often gives them the opportunity to work in technical areas that are completely new to them.

**Acknowledgements**
We thank the 64 students who have taken the module for their creativity, innovation and enthusiasm.

**Appendix A (Bill of Materials)**

| Item | Approx Cost |
|---|---|
| **Lab Materials** | |
| **Hardware** | **€535** |
| *Raspberry Pi B+ (6 units)* | *€180* |
| *uBlox-7 GPS (4 units)* | *€60* |
| *5000mAh Power Pack (6 units)* | *€120* |
| *Onboard Sensor Suite (incl. internal & external DS18B20 temperature sensors, MS5611 pressure sensors)* | *€75* |
| *Other Electronics (incl., wires & components, breadboards, tools)* | *€100* |
| **T3 TupperSat Telemetry Thingamabob (6 units)** | **€100** |
| *RFM95W LoRa Radio (6 units)* | *€80* |
| *ATMega328 Microcontroller (6 units)* | *€10* |
| *PCBs* | *€10* |
| **Student Payload Budget (€100 per team)** | **€400** |
| **Launch Campaign** | |
| **Materials** | **€790** |
| *Balloons* | *€180* |
| *Parachutes* | *€180* |
| *Compressed Helium* | *€350* |
| *Extras (incl., rope, tape, cable ties, sheeting)* | *€40* |
| **Transportation** | **€210** |
| *Vehicle Hire* | *€120* |
| *Fuel Costs* | *€90* |